\documentclass[prd,showpacs,secnumarabic,12pt]{revtex4}
\usepackage{amsmath}

\newcommand{\pni}{\par\noindent}
\begin{document}
\title{Brane cosmology with $^{(4)}R$ term in the bulk} 
\author {M. La Camera} \email{lacamera@ge.infn.it} 
\affiliation {Department of Physics and INFN - University of 
Genoa\\Via Dodecaneso 33, 16146 Genoa, Italy} 
\begin{abstract}
We consider brane cosmology when the 4D Ricci scalar term is 
added to the 5D Einstein-Hilbert action and discuss the 
role that the addition of this term has on  the brane-bulk 
system. The induced brane dynamics is shown to be the usual 
Einstein dynamics coupled to a modified energy-momentum tensor 
which is well defined once the 5D Einstein equations are solved 
in the bulk.  The 5D Einstein equations valid everywhere 
in the bulk, but not in the brane, are projected on the brane. 
Then making use for the embedding of the brane in the bulk of the
Israel junction conditions, modified by a source term coming from
the addition of the intrinsic curvature scalar in the bulk 
action, it is possible to obtain the effective 4D Einstein 
equations on the brane consistent with the bulk geometry. 
\end{abstract} 
\maketitle  
\section{I\lowercase{ntroduction}}
In the present work we investigate the cosmological evolution of 
the brane-bulk system in the framework of the Randall-Sundrum 
$(A)dS_5$ scenario [1]. The effective 4D gravitational  equations
in the brane without curvature correction terms were first 
obtained by Shiromizu, Maeda and Sasaki [2]. These equations 
have been later recovered and generalized both on the brane and 
in the bulk taking into account the  effect of a general bulk 
energy-momentum tensor and either the asymmetric embedding [3] or
the accelerations of normals [4]. However, even employing  more 
generalized gravitational actions, the derived 4D Einstein 
equations do not in general form a closed system due to the 
presence of a Weyl term which can only be specified in terms of 
the bulk metric, so other equations are to be written down and 
different procedures arise in splitting the non-Einsteinian terms
between bulk and brane [5]. We assume omogeneity and isotropy in 
the three ordinary spatial dimensions but these symmetries cannot
be extended to the extra dimension due to the presence of the 
brane, so all the physical quantities will depend on time and on 
the extra dimension. The solutions of the 5D Einstein equations 
that we shall obtain will be valid strictly in the bulk.  Then we
project on the brane at $y=0$ those equations and making use of 
the junction conditions, modified by an additional term coming 
from the $^{(4)}R$ curvature correction, as the boundary 
conditions imposed for the embedding of the brane in the bulk
we obtain the effective 4D Einstein  equations on the brane 
consistent with the bulk geometry. The method used here of 
deriving the 5D solution that is then projected  onto the brane 
to study the brane dynamics was used extensively in the early 
days of braneworld cosmology by P.Kanti et al [6]. The  paper 
is organised as follows. In the next Section, before giving a 
brief account of our method, we summarize the results  obtained 
by Kofinas [7] when the $^{(4)}R$ term is included in the 5D 
action. Such a term, that was considered by a number of authors 
in the literature (see [8],[9],[10] and references therein), is 
generically introduced by quantum corrections coming from the 
bulk gravity and its coupling with matter confined to the brane, 
moreover its inclusion brings a convenient decomposition of the 
matter terms.  In Section 3 we find  the related  equations  in 
the brane, assumed infinitely thin and $\mathbf{Z_2}$ symmetric 
in the bulk. In Section 4 we use the flexibility of the 5D 
solution to describe some cosmological models in 4D.  Finally,  
in the Appendix we show how the 5D dynamical solution we adopt 
can be obtained starting from a 5D static solution.\pni 
\textit{Conventions}. Throughout  the  paper the 5D metric 
signature is taken to be $(+,+,+,-,\varepsilon)$ where 
$\varepsilon$ can be $+1$ or $-1$ depending on whether the extra 
dimension is spacelike or timelike, while the choice of the 4D 
metric signature is $(+,+,+,-)$. The spacetimes coordinates are 
labelled $x^i=(r,\vartheta,\varphi), x^4=t$. The extra coordinate
is $x^5=y$. Bulk indices will be denoted by capital Latin letters
and brane indices by lower Greek letters. In what follows we 
choose units such that $\hbar = c = 1$. 
\section{B\lowercase{raneworld} E\lowercase{instein field 
equations}} In this section we recall the results obtained by 
Kofinas [7], which we shall use in the following, giving a brief 
account of their derivation. Once we have solved the equations in
the bulk, the form of the induced equations will allow us finding
brane solutions following the methods of General Relativity with 
a well defined energy-momentum tensor. The starting point in [7] 
is a three-dimensional brane $\Sigma$ embedded in a 
five-dimensional spacetime $M$. For convenience the coordinate 
$y$ is chosen such that the hypersurface $y = 0$ coincides with 
the brane. The total action for the system is taken to be 
\begin{eqnarray} \mathcal{S} && = \dfrac{1}{2\,\kappa_{5}^2}\, 
\int_M\, \sqrt{-\varepsilon\,^{(5)}g} \,(^{(5)}R - 
2\Lambda_5)\,d^5x + \dfrac{1}{2\,\kappa_{4}^2} \,\int_\Sigma\, 
\sqrt{-\,^{(4)}g} \,(^{(4)}R - 2\Lambda_4)\,d^4x \nonumber\\ && 
\,\, +  \int_M\, \sqrt{-\varepsilon\,^{(5)}g} \,L_5^{mat}\,d^5x +
\int_\Sigma\, \sqrt{-\,^{(4)}g} \,L_4^{mat}\,d^4x 
\end{eqnarray} 
The constants $\kappa_5^2$ and $\kappa_4^2$ are given by 
\begin{equation}
\kappa_5^2 = 8\pi G_5 = M_5^{-3} \quad,\quad  \kappa_4^2 = 8\pi 
G_4 = M_4^{-2}
\end{equation}
where  $M_5$ and $M_4$ are the Planck masses. 
Varying (1) with respect to the  
bulk metric $g_{AB}$ one obtains the equations
\begin{equation}
^{(5)}G_A^B = - \Lambda_5 \delta_A^B + \kappa_5^2\, (^{(5)}T_A^B 
+ ^{(loc)}T_A^B \delta(y)) 
\end{equation}
where
\begin{equation}
^{(loc)} T_A^B = -\, 
\dfrac{1}{\kappa_4^2}\,\sqrt{\dfrac{-\,^{(4)}g}{-\varepsilon\,^{(5)}g}}\, 
(^{(4)}G_A^B - \kappa_4^2\,{^{(4)}}T_A^B + \Lambda_4h_A^B)
\end{equation}
is the localized energy-momentum tensor of the brane. 
$^{(5)}G_{AB}$ and $^{(4)}G_{AB}$ denote the Einstein tensors 
constructed from the bulk and the brane metrics respectively, 
while the tensor $h_{AB} = g_{AB} -\varepsilon n_An_B$ is the 
induced metric on the hypersurfaces $y$ = constant, with $n^A$ 
the normal unit vector on these
\begin{equation}
n^A = \dfrac{\delta_5^A}{\Phi}, \hspace{0.5in} n_A = 
(0,0,0,0,\varepsilon \,\Phi) 
\end{equation}
The scalar $\Phi$ which normalizes $n^A$ is known as the lapse 
function and in a cosmological scenario which we shall consider 
later, it will depend on $t$ and $y$.  The way the  coordinate 
$y$ has been chosen allows to write the five-dimensional line 
element, at least in the neighborood of the brane, as 
\begin{equation}
dS^2 = g_{AB}\, dx^A dx^B = g_{\mu\nu}\, dx^\mu dx^\nu +\varepsilon 
\,\Phi^2dy^2 
\end{equation}
Using the methods of   canonical analysis [11] the Einstein eqs. 
(3) in the bulk  are split into the following sets of equations 
\begin{subequations}
\begin{eqnarray} 
K_{\mu;\nu}^\nu -\, K_{;\mu}  &&=\, \varepsilon\, \kappa_5^2\, 
\Phi\,^{(5)}T_\mu^y \label{equation20}\\
K_\nu^\mu K_\mu^\nu - K^2 +\, \varepsilon\,{^{(4)}R}  &&=\, 
 2\, \varepsilon \,(\Lambda_5 -\, \kappa_5^2\,{^{(5)}}T_y^y) 
\label{equation21}\\ \hspace{-0.3in} \dfrac{\partial 
K_\nu^\mu}{\partial y} + \Phi\,K\,K_\nu^\mu -\, \varepsilon\, 
\Phi\,^{(4)}R_\nu^\mu + \varepsilon\, 
g^{\mu\lambda}\,\Phi_{;\lambda\nu} &&=\, -\,\varepsilon 
\kappa_5^2\,\Phi\,\left(^{(loc)}T_\nu^\mu -\, 
\dfrac{1}{3}\,^{(loc)} T\,\delta_\nu^\mu\right)\,\delta(y) 
\nonumber\\  &&\quad\, -\, \varepsilon\, 
\kappa_5^2\,\Phi\,^{(5)}T_\nu^\mu 
+\dfrac{\varepsilon}{3}\,\Phi\,(\kappa_5^2\,{^{(5)}T} -\, 
2\Lambda_5)\,\delta_\nu^\mu \label{equation22} 
\end{eqnarray} 
\end{subequations}
where $K_{\mu\nu}$ is the extrinsic curvature of the 
hypersurfaces $y$ = constant:
\begin{equation} 
K_{\mu\nu} = \dfrac{1}{2\,\Phi}\,\dfrac{\partial 
g_{\mu\nu}}{\partial y}, \quad K_{A y} = 0 
\end{equation} 
The Israel junction conditions [12] for the singular part 
in eq. (7c) are 
\begin{equation}
[K_\nu^\mu] = -\varepsilon\, 
\kappa_5^2\,\Phi_0\,\left(^{(loc)}T_\nu^\mu -\, 
\dfrac{1}{3}\,^{(loc)} T\,\delta_\nu^\mu\right) 
\end{equation}
where the square brackets mean discontinuity of the quantity 
across $y = 0$ and $\Phi_0$ represents $\Phi$ at $y=0$. 
Consequently, considering a $\mathbf{Z_2}$ symmetry on reflection
around the brane, (9) becomes
\begin{equation}
^{(4)}G_\nu^\mu = - \Lambda_4\,\delta_\nu^\mu + \kappa_4^2\, 
{^{(4)}T_\nu^\mu}  + \dfrac{2 \varepsilon}{r_c}\,(\overline{K}_\nu^\mu
- \overline{K}\delta_\nu^\mu)
\end{equation}
where \,$\overline{K}_\nu^\mu = K_\nu^\mu(y=0^+) = 
-\,K_\nu^\mu(y=0^-)$\, and \,$r_c = 
\kappa_5^2/\kappa_4^2$\, is a crossover term which determines 
the region of validity of conventional  four-dimensional General 
Relativity. From eq. (7a) it follows that the tensor  
$^{(4)}T_\nu^\mu$ satisfies the  conservation 
law $^{(4)}{T_\nu^\mu}_{;\mu} = 0$ 
provided $^{(5)}T_\mu^y =0$, which means no exchange of energy 
between brane and bulk. The quantities $\overline{K}_\nu^\mu$ are
still undetermined and should be obtained from some exact 
solution of the global five-dimensional spacetime. To determine 
the equations on the brane one can follow the method suggested 
in ref. [2], but this will reveal a difficult task due to the 
necessity of taking into account the evolution of the Weyl term 
to close the system of equations. A different approach, as 
discussed by Bin$\grave{e}$truy et al. [13], is to solve the 5D 
Einstein equations strictly in the bulk ($y\neq 0$) and then to 
take the brane into account by using the Israel junction 
conditions. In this work we shall keep in mind this latter 
approach but, having added the $^{(4)}R$ term, we shall make  a 
different use of the junction conditions. More in detail, we 
start projecting on the brane the  solution obtained in the bulk 
without considering the  distributional part at $y=0$. This will 
be done using the geometrical identity 
\begin{equation} 
^{(4)}R_{BCD}^A =\, 
^{(5)}R_{NKL}^M\,h_M^A\,h_B^N\,h_C^K\,h_D^L + 
\varepsilon \, (K_C^A\,K_{BD} - K_D^A\,K_{BC}) 
\end{equation} 
and taking suitable contractions from  the above relation. So it 
is possible to construct the four and five-dimensional Einstein 
tensors and to get finally the parallel to the brane equations 
\begin{eqnarray}
^{(4)}G_\nu^\mu = && -\,\dfrac{1}{2}\,\Lambda_5 \delta_\nu^\mu 
+ \dfrac{2}{3}\,\kappa_5^2\, \left({}^{(5)}\overline{T}_\nu^\mu +
\left({}^{(5)}\overline{T}_y^y -  
\dfrac{1}{4}\,{}^{(5)}\overline{T}\right) \delta_\nu^\mu  \right)
\nonumber \\ 
&& +\varepsilon \, \left(\overline{K}\,\overline{K}_\nu^\mu - 
\overline{K}_\lambda^\mu\, \overline{K}_\nu^\lambda \right) + 
\dfrac{\varepsilon}{2}\, \left(\overline{K}_\lambda^\kappa\, 
\overline{K}_\kappa^\lambda - \overline{K}^2\right) 
\delta_\nu^\mu - g^{\kappa\mu}\,{}^{(5)}\overline{C}_{\kappa y 
\nu}^y 
\end{eqnarray}
Here ${}^{(5)}C_{\kappa y \nu}^y$ is the ``electric'' part of 
the bulk Weyl tensor, while $\overline{T}$ and $\overline{C}$ are
the limiting values of those quantities at $y = 0^+$ or $0^-$.
Once we have solved the Einstein equations strictly in the bulk 
we can make explicit the various terms appearing in the 
right-hand side of (12). Now we impose to the above solution 
boundary conditions to take into account the physical presence of
the brane. This can be done if we consider equations  (10) as the
boundary conditions imposed for the embedding of the brane in the
bulk so  we  have to  equate the two independent equations (10) 
and  (12). In this way, however, we would obtain the ``bare'' 
quantities $\Lambda_4$ and ${^{(4)}T_\nu^\mu}$ but not  the 
effective quantities as seen by an observer  confined to the 
brane. For a brane observer eqs. (10) are instead written as the 
usual Einstein equations 
\begin{equation} 
^{(4)}G_\nu^\mu = - \Lambda_{4,eff}\,\delta_\nu^\mu + \kappa_4^2\, 
\left({^{(4)}T_\nu^\mu}\right)_{eff} 
\end{equation}
So  the cosmological constant  $\Lambda_{4,eff}$ and the 
effective energy-momentum tensors  
$\left({^{(4)}T^\mu_\nu}\right)_{eff}$ can be obtained 
equating the right-hand sides of eqs. (12) and (13). 
\section{D\lowercase{ynamics in the brane-bulk system}} We 
consider the 5D metric in the form commonly used in cosmological 
applications 
\begin{equation} dS^2 = a^2(t,y)\,d\sigma_k^2 - 
n^2(t,y)\,dt^2 + \varepsilon\,\Phi^2(t,y)\,dy^2
\end{equation}
where 
\begin{equation}
d\sigma_k^2 = \dfrac{dr^2}{1-k r^2} + r^2  
(d\vartheta^2+\sin^2\vartheta d\varphi^2)
\end{equation}
and $k = +1, 0, -1$ is the curvature index. Having specified the 
form of the metric, we now turn to the 5D Einstein equations (3) 
considered strictly in the bulk, that is, without the   
energy-momentum tensor at $y=0$. These equations  can be solved 
once the structure and the content of the bulk come as a result 
of a physically acceptable theory in higher dimensions. Exact 
time-dependent solutions which generalize the static solutions 
were constructed using diffeomorphism invariance [14,15]. 
Kehagias and Tamvakis [16] transformed the static Randall-Sundrum
$(A)dS_5$ solution  into a dynamical one by considering boosts 
along the fifth dimension and  found a time-dependent 5D solution
for a bulk with vacuum energy but otherwise empty and with 
vanishing Weyl tensor. In the  present work we want instead to 
consider (see (12)) a bulk where the non-localized 
energy-momentum tensor and the ``electric'' part of the Weyl 
tensor are different from zero, so we should start by a well
defined bulk matter content described by the tensor  
$^{(5)}T_A^B$ and then solve the field equations. However
our aim is to overcome  the problem  of the brane field equations
being non-closed so, to give an illustrative example of our 
method, we shall proceed in a bit unhorthodox way. In 
order to have a simple and non-trivial dynamical 5D solution we 
start from a static Randall-Sundrum $(A)dS_5$ bulk and we
construct, generalizing the transformations in [16] a dynamical 
5D line element with non-vanishing 5D Weyl tensor. Subsequently 
we obtain the correspondent energy-momentum content using the 
Einstein equations. This manner of proceeding  may be justified 
by the fact that this work is mainly focused on the brane 
phenomenology of the model. Here we anticipate the main results 
of the procedure and defer to the Appendix for detailed 
calculations.  Our dynamical line element will be obtained 
transforming the static Randall-Sundrum $(A)dS_5$ metric, where 
the three-space is not necessarily flat but has a curvature index
$k = +1, 0, -1$, into a dynamical one by considering boosts along
the fifth dimension. Then we take into account the Einstein  
equations in the bulk, away from the brane at $y=0$,  and 
require that there is no energy flow from the brane towards the
bulk and vice-versa, which implies ${}^{(5)}G_t^y = 0$. The 
above constraint is easily satisfied if one  chooses wave-like 
expression for the metric coefficients so, assuming the $Z_2$ 
symmetry $y \to -\,y$, it follows that  $a(t,y)$, $n(t,y)$ and 
$\Phi(t,y)$ in the line element (14) will be function of 
$w=t-\,\lambda\, |y|$ with $\lambda$ a dimensionless constant. 
Metric coefficients in the form of plane waves propagating in the
fifth dimension have previously been used in the literature in 
somewhat different contexts [17,18,19,20]. Finally, the bulk line
element away from the brane was found to depend only by the scale
factor $a(t,y) = a(t-\,\lambda\, |y|)$ in the form 
\begin{eqnarray} 
dS^2 = a^2\, d\sigma_k^2 
\hspace{-.3cm}  &&-\, \dfrac{1}{2\,\gamma^2 \lambda^2}\, 
\left(\kappa^2 a^2 + \sqrt{\kappa ^4 a^4 + 4\,\varepsilon\, 
\gamma^2 \lambda^2\, ({\overset{\ast}{a}})^2}\right)\, dt^2 
\nonumber \\ &&+\, \dfrac{1}{2\,\gamma^2}\, \left(-\, \kappa^2 
a^2 + \sqrt{\kappa ^4 a^4 + 4\,\varepsilon\, \gamma^2 \lambda^2\,
({\overset{\ast}{a}})^2}\right)\, dy^2 
\end{eqnarray}
where $\kappa$ is the constant scale factor for the extra 
dimension, the superscribed asterisk $\overset{\ast}{\frac{}{}}$ 
denotes derivative with respect to $w$ and $\gamma$ is a 
dimensionless constant which comes from the constraint 
${}^{(5)}G_t^y = 0$, namely ${\overset{\ast}{a}} = 
\gamma\,n\,\Phi$. It should be noted that $a(t - \lambda \,  
|y|)=0$ corresponds to a scale factor singularity for the 5D 
model which is similar to those that occur in the 4D 
Friedmann-Robertson-Walker models. From the following 
curvature invariants 
\begin{equation} 
\hspace{-0.5cm}^{(5)}R=-20\epsilon\kappa^2+
\dfrac{6k}{a^2(t-\lambda \, |y|)}\hspace{0.5cm} 
^{(5)}R_{AB}{^{(5)}R^{AB}}=80\epsilon\kappa^4-\dfrac{48\epsilon k
\kappa^2}{a^2(t-\lambda \,|y|)}+\dfrac{12k^2}{a^4(t-\lambda \,|y|)}
\end{equation}
\begin{equation}
^{(5)}R_{ABCD}{^{(5)}R^{ABCD}}=40\epsilon\kappa^4-\dfrac{24\epsilon
k\kappa^2}{a^2(t-\lambda \,|y|)}+\dfrac{12k^2}{a^4(t-\lambda 
\,|y|)}
\end{equation}
we see that there is no other singularity except the  one which 
may unaivodably occur if the scale factor of the fifth dimension 
vanishes in some $(t,y)$ hyperplane.\pni Now we calculate the
5D Einstein tensor and from the field equations  considered 
strictly in the bulk we obtain the cosmological constant 
$\Lambda_5$  and the  energy-momentum tensor $^{(5)}T_A^B$ as
\begin{equation} 
\Lambda_5 = -\,6\,\varepsilon\,\kappa^2 , \qquad ^{(5)}T_A^B = 
\textrm{diag}(p_B,p_B, p_B,-\,\rho_B,p_\perp) 
\end{equation} 
where 
\begin{equation} \kappa_5^2\,p_B = 
-\,\dfrac{k}{a^2(t-\lambda \,|y|)},  \quad \kappa_5^2\,\rho_B = 
-\,\kappa_5^2\,p_\perp = \dfrac{3\,k}{a^2(t-\lambda \,|y|)} 
\end{equation}
the subscript ${}_B$ referring to the bulk. A comment is needed 
about the cosmological fluid described by the energy-momentum 
tensor which arises  from the curvature index $k$ in eq. (15).
It obeys an equation of state $p_B = (\gamma - 1)\,\rho_B$ with 
barotropic index $\gamma = 2/3$, its pressure and energy density 
are proportional to $k$ and scale as $a^{-2}$. All these features
lead to the interesting possibility that the energy-tensor (20) 
can describe a fluid composed of cosmic strings, as discussed by 
a number of authors [21].  We can 
use the flexibility  of the metric (16) to choose many different 
5D scale factors, but clearly each choice must meet the necessary
requirements to give models acceptable on physical grounds.\pni 
Let us now deal with the brane dynamics. We consider homogeneous 
and isotropic geometries in the brane so the effective tensor 
$\left({^{(4)}T^\mu_\nu}\right)_{eff}$ will describe a 
cosmological fluid endowed with pressure $p_{eff}$ and energy 
density $\rho_{eff}$.  Equating eqs. (12) and (13)  we obtain 
\begin{eqnarray}
-\, \Lambda_{4,eff} + 
\kappa_4^2\, p_{eff} &=& -\,\dfrac{\Lambda_5}{2} -  
\dfrac{\epsilon}{\Phi_0^2} \left(\dfrac{1}{a}\dfrac{\partial 
a}{\partial y}\right)_0 \left(\dfrac{1}{a}\dfrac{\partial 
a}{\partial y} + \dfrac{2}{n}\dfrac{\partial n}{\partial 
y}\right)_0 - \dfrac{k}{a_0^2} \\
-\, \Lambda_{4,eff}
-\,\kappa_4^2 \,\rho_{eff} &=& -\,\dfrac{\Lambda_5}{2}  -\, 
\dfrac{3\,\epsilon}{\Phi_0^2} \left(\dfrac{1}{a}\dfrac{\partial 
a}{\partial y}\right)_0^2 -\,\dfrac{3 k}{a_0^2}
\end{eqnarray}
The effective  cosmological constant  $\Lambda_{4,eff}$ may 
be different from $\Lambda_5/2 = -3\,\epsilon\,\kappa^2$, its 
value being modified by possible additive constant   terms 
contained in (21) and (22). The Einstein tensor $^{(4)}G_\nu^\mu$
which appears in the left-hand sides of (12) and (13) is 
constructed from the brane metrics 
\begin{equation} 
ds^2 = \widetilde{a}^2(t)\,d\sigma_k^2 - \widetilde{n}^2(t)\, 
dt^2 \end{equation} 
Now, in higher-dimensional theories there is the question of 
which metric frame is the correct representation of our 
four-dimensional spacetime. In  many braneworld theories, the 
physical metric in 4D is identified with the induced one, while 
in other approaches the physical metric either is assumed to be 
conformally related to the induced one or is determined by the 
condition of classical confinement in the absence of 
non-gravitational forces [22]. For the sake of simplicity, here 
we choose to identify  the metric (23) with the induced one, so
we have $\widetilde{a}(t) = a(t,0) \equiv a_0(t)$  and 
$\widetilde{n}(t) = n(t,0)\equiv n_0(t)$. It follows that eqs. 
(12) and (13) are identically satisfied by the Einstein tensor 
$^{(4)}G_\nu^\mu$ constructed from (23). From the knowledge 
$a_0(t)$ and $n_0(t)$ one can also obtain other cosmological 
quantities such as the Hubble parameter $H=\dot{a_0}/(n_0 a_0)$ 
or the deceleration parameter $q = -\, 
(a_0\,\ddot{a_0})/\dot{a_0}^2 + (a_0\, 
\dot{n_0})/(\dot{a_0}\,n_0)$. Difficulties may instead arise from
the exact evaluation of the 4D proper time $\tau$ when dealing 
with  the integral $\tau = \displaystyle{\int} 
n_0\,dt$ and a generic value of $n_0(t)$. Finally, if we define 
\begin{equation} 
\kappa_4^2\,p_\phi = \kappa_4^2\,p_{eff} + 
\dfrac{k}{a_0^2} \quad \mathrm{and} \quad \kappa_4^2\,\rho_\phi =
\kappa_4^2\,\rho_{eff} - \dfrac{3 k}{a_0^2} 
\end{equation} 
we can model the fluid  in terms of a scalar field $\phi$, 
minimally coupled to Einstein gravity and self-interacting 
through a potential $V(\phi)$, with pressure and energy density 
given by \begin{eqnarray} p_\phi & = &  \pm\,\dot{\phi_0}^2/(2 
n_0^2) - V \\ \rho_\phi & = & \pm\, \dot{\phi_0}^2/(2 n_0^2) + V 
\end{eqnarray}
where the upper (lower) sign corresponds to a standard (phantom) 
scalar field.
\section{S\lowercase{ome possible brane scenarios}}
In this section we describe two of the possible brane scenarios  
consistent with our bulk solution. We shall  choose 
simple values for the scale factor $a(t,y)$ and then determine 
$\Lambda_{4,eff}$ and $\left({^{(4)}T^\mu_\nu}\right)_{eff}$ 
together with the parameters $q$ and $H$.  In the following we  
shall consider a spacelike fifth dimension, so $\epsilon=1$ and 
give  relevant 4D quantities as a fuction of the 4D  proper time 
$\tau$. \pni 
\textbf{A)}\,  Let us first  consider 
the case $a(t,y) = (\gamma \lambda/\kappa)\, \sin{\kappa 
(t-\lambda |y|)}$.\pni This choice gives $n_0(t) = 1$ so the 
coordinate time $t$ now coincides with the 4D proper time $\tau$ 
and therefore \begin{equation} 
a_0(\tau)= (\gamma \lambda/\kappa)\, \sin{\kappa\,\tau}, \quad
q(\tau)=\tan^2{\kappa\, \tau}, \quad  
H(\tau)=\kappa\,\cot{\kappa\,\tau} \end{equation}
The evolution of the universe begins with a big bang at $\tau = 
0$, reaches a maximum $\left({a_0}\right)_{max} = 
(\gamma\,\lambda)/\kappa$ and terminates with a big rip at 
$\kappa\,\tau =\pi$. We do not give numerical values for $\gamma$
and $\lambda$ while $\kappa$ depends on the scale factor of the 
fifth dimension. The cosmological constant  $\Lambda_{4,eff}$ and
the tensor $\left({^{(4)}T^\mu_\nu}\right)_{eff}$ are given by 
\begin{equation} 
\Lambda_{4,eff} = -\,3\,\kappa^2, \quad 
p_{eff} = -\, \dfrac{(\gamma^2\,\lambda^2+k)}{\kappa_4^2\, 
a_0^2(\tau)},\quad  
\rho_{eff} = \dfrac{3\,(\gamma^2\,\lambda^2+k)}{\kappa_4^2\, 
a_0^2(\tau)} 
\end{equation}
As to the standard scalar field, we have:
\begin{equation}
\phi = \dfrac{\sqrt{2}}{\kappa_4}\,\ln{\tan{\frac{\kappa\, 
\tau}{2}}}, \quad 
V = \dfrac{2\,\kappa^2}{\kappa_4^2 \sin^2{\kappa\,\tau}}, \quad 
V(\phi)=  \dfrac{2\,\kappa^2}{\kappa_4^2}\,  
\cosh^2{\dfrac{\kappa_4\,\phi}{\sqrt{2}}}
\end{equation} 
The same results, starting from different points 
of view, were obtained in [23]. \pni
\textbf{B)}\, Now let us consider the case $a(t,y) = (\gamma 
\lambda /\kappa)\,(1-\kappa (t -\lambda |y|)^{-\,1}$.\pni 
This choice gives $n_0(t) = \sqrt{2/(\sqrt{5}-\,1)}\, 
(1-\kappa\,t)^{-1}$ so the relation between the 
coordinate time $t$ and the 4D proper time $\tau$ is 
$(1-\kappa\,t) = \exp{\left[-\,  
\sqrt{(\sqrt{5}-\,1)/2}\,\,\kappa\,\tau\right]}$. Therefore
\begin{equation}  
a_0(\tau) = \frac{\gamma \,\lambda}{\kappa}  \,\exp{\left(\sqrt 
{\dfrac{\sqrt{5}-1}{2}}\,\kappa\, \tau\right)},\quad 
q=-\,1, \quad H = \sqrt{\dfrac{\sqrt{5}-1}{2}}\,\kappa 
\end{equation}
The cosmological constant  $\Lambda_{4,eff}$ and
the tensor $\left({^{(4)}T^\mu_\nu}\right)_{eff}$ are given by 
\begin{equation}
\Lambda_{4,eff} = \dfrac{3\,(\sqrt{5}-1)}{2}\,\,\kappa^2, 
\quad p_{eff} = -\, \dfrac{k}{\kappa_4^2\, a_0^2(\tau)},\quad  
\rho_{eff} = \dfrac{3\,k}{\kappa_4^2\, a_0^2(\tau)} 
\end{equation}
The values of  $a(t,y)$ chosen in the previous illustrative 
examples reproduce results already known in the literature.
Less simple choices for the 5D scale factor may describe new  
brane scenarios but also may require a more involved treatment. 
In conclusion, this paper investigates the influence of the 
$^{(4)}R$ term included in the bulk action on the spherically 
symmetric braneworld solutions. The brane dynamics is made closed
by using the modified junction conditions as the boundary 
conditions for the embedding of the brane in the bulk, so it is 
possible to obtain brane cosmological solutions consistent with 
the bulk geometry. We started from a particularly simple 
time-dependent solution in the bulk away from the brane, but 
other physically acceptable solutions in the bulk can be 
considered provided that the related brane dynamics is in 
accordance to the observations on the brane. 
\appendix*
\section{T\lowercase{ransforming  static bulk  solutions into 
dynamical ones}} 
The Randall-Sundrum $(A)dS_5$  model  is the simple  braneworld 
with curved extra dimension that allows for a meaningfull 
approach to cosmology, therefore we start from this model but, at
this point, we do not yet require $Z_2$ symmetry on reflection 
around the value $Y=0$ so we write 
\begin{equation} 
dS^2 = e^{-2\, \kappa \,Y}\,\left( A^2\, d\sigma_k^2
-dT^2 \right)+ \varepsilon \,dY^2 
\end{equation}
Here $\kappa$ and $A$ are, respectively, the constant scale 
factors for the extra dimension $Y$ and for the ordinary 
three-space and $d\sigma_k^2$ is the line element of maximally 
symmetric three-spaces with curvature index $k = +1, 0, -1$:
\begin{equation}
d\sigma_k^2 = \dfrac{dr^2}{1-k r^2} + r^2  
(d\vartheta^2+\sin^2\vartheta d\varphi^2)
\end{equation}
Since  our purpose is to describe the time evolution on the 
braneworld,  we need to transform the static bulk solution (A.1) 
into a dynamical one. This goal was already achieved in 
literature where dynamical  solutions are derived from the static
Randall-Sundrum $(A)dS_5$ metric by considering boosts along the 
fifth dimension [16]. Applied to the actual case, we 
generalize those transformations  as 
\begin{equation} 
\begin{cases}
T = & \dfrac{\dfrac{1- F(t,y)}{\chi}- \varepsilon \,\dfrac{\chi} 
{\kappa^2}G(t,y)}{\sqrt{1- \,\dfrac{\chi^2}{\kappa^2}}} 
\\ {}& \\ e^{\kappa\, Y} = & \dfrac{F(t,y)+G(t,y)-1} 
{\sqrt{1-\dfrac{\chi^2}{\kappa^2}}} 
\end{cases} 
\end{equation} 
where $F(t,y)$ and $G(t,y)$ are dimensionless functions
and $\chi$ is a constant, with the dimensions of $\kappa$, 
related to the boost along the fifth dimension. The coordinate 
$y$ is chosen so that the hypersurface $y=0$ coincides with the 
brane. \pni As a result the metric (A.1) becomes:
\begin{equation}
\hspace{-0.1cm} dS^2 = 
\dfrac{1}{(F+G-1)^2}\left\{\left(\dfrac{\kappa^2-\chi^2}{\kappa^2}\right) 
\, A^2 \, d\sigma_k^2 +\left(\dfrac{\kappa^2-\varepsilon\, 
\chi^2}{\kappa^4\,\chi^2}\right)\, \left[-\,\kappa^2\, (dF)^2 
+\varepsilon\, \chi^2 \, (dG)^2\right]\right\}
\end{equation}
Note that the static line element (A.1) can be recovered from the
above equations on condition that as $\chi \to 0$ it  
results $F \approx 1-\chi\, t - \varepsilon\,  (\chi^2/\kappa^2) 
\,e^{\kappa\, y}$ and $G \approx \chi\, t + 
e^{\kappa \, y}$. \pni 
The line element (A.4)  is in the form   commonly  used in 
cosmological applications 
\begin{equation}
dS^2 = a^2(t,y)\,d\sigma_k^2 - n^2(t,y)\,dt^2 +  
\varepsilon\,\Phi^2(t,y)\,dy^2
\end{equation}
Now we can choose suitable functions $F$  and $G$ to  obtain 
explicit expressions for the metric coefficients $a$, $n$ and 
$\Phi$. Comparing eqs. (A.4) and (A.5) we get
\begin{subequations}
\begin{eqnarray}
\dfrac{A^2\,(\kappa^2-\chi^2)}{\kappa^2\,(F+G-1)^2} = && a^2 
\label{equation30}\\
\dfrac{a^2\,(\kappa^2-\varepsilon \, \chi^2)}{A^2\, \kappa^2\, 
\chi^2 \, (\kappa^2-\chi^2)}\,\left[\kappa^2\,\left(\dfrac{\partial 
F}{\partial t}\right)^2 -\, \varepsilon \, \chi^2 
\left(\dfrac{\partial G}{\partial t}\right)^2\right] = && n^2 
\label{equation31}\\ 
-\,\dfrac{a^2\,(\kappa^2-\varepsilon \, \chi^2)}{A^2\, \kappa^2\,
\chi^2 \, 
(\kappa^2-\chi^2)}\,\,\left[\kappa^2\,\left(\dfrac{\partial 
F}{\partial y}\right)^2 -\, \varepsilon \, \chi^2 
\left(\dfrac{\partial G}{\partial y}\right)^2\right] = && 
\varepsilon\,\Phi^2 \label{equation32}\\ 
\kappa^2\,\left(\dfrac{\partial F}{\partial 
t}\right)\,\left(\dfrac{\partial F}{\partial y}\right) -\, 
\varepsilon \, \chi^2 \left(\dfrac{\partial G}{\partial 
t}\right)\,\left(\dfrac{\partial G}{\partial y}\right)= && 0 
\label{equation33}
\end{eqnarray}
\end{subequations}
  Once the new metric coefficients are known
it is possible from the  Einstein equations (3)  in the bulk, 
that is, away from the brane at $y=0$, to obtain the 
energy-momentum tensor ${}^{(5)}T_A^B$. This can be 
achieved by recalling that in the coordinate system (A.5) 
the non-vanishing components of the Einstein tensor 
$G_A^B$ are 
\begin{subequations} \begin{eqnarray}
\hspace{-1in} G_r^r = G_\vartheta^\vartheta = G_\varphi^\varphi =
&&-\,\dfrac{1}{n^2}\left[\dfrac{\ddot{\Phi}}{\Phi} + \dfrac{2 
\ddot{a}}{a} + \dfrac{\dot{\Phi}}{\Phi}\,
\left(\dfrac{2\dot{a}}{a} - \dfrac{\dot{n}}{n}\right) + 
\dfrac{\dot{a}}{a}\,\left(\dfrac{\dot{a}}{a} - \dfrac{2 
\dot{n}}{n}\right)\right]  \nonumber \\
&&  +\,\dfrac{\varepsilon}{\Phi^2}\,\left[\dfrac{2a''}{a} + 
\dfrac{n''}{n} +\dfrac{a'}{a}\,\left(\frac{a'}{a} + 
\dfrac{2n'}{n}\right) - \dfrac{\Phi'}{\Phi}\,\left(\dfrac{2a'}{a}
+ \dfrac{n'}{n}\right)\right] - \dfrac{k}{a^2} 
\label{equationc}\\
G_t^t = &&  -\,\dfrac{3}{n^2}\,\left(\dfrac{\dot{a}^2}{a^2}+ 
\dfrac{\dot{a}\dot{\Phi}}{a\Phi}\right) + 
\dfrac{3\,\varepsilon}{\Phi^2}\, \left(\dfrac{a''}{a} + 
\dfrac{{a'}^2}{a^2} - \dfrac{a'\Phi'}{a\Phi}\right) - 
\dfrac{3k}{a^2} \label{equationd}\\
G_y^y =&&   -\,\dfrac{3}{n^2}\,\left(\dfrac{\ddot{a}}{a} + 
\dfrac{\dot{a}^2}{a^2} - \dfrac{\dot{a}\dot{n}}{an}\right) + 
\dfrac{3\,\varepsilon}{\Phi^2}\,\left(\dfrac{{a'}^2}{a^2} + 
\dfrac{a'n'}{an}\right) - \dfrac{3k}{a^2} \label{equatione}\\
G_t^y =&& 
 -\,\dfrac{3\,\varepsilon}{\Phi^2}\,\left(\dfrac{\dot{a}'}{a} - 
\dfrac{\dot{a}n'}{an} - \dfrac{a'\dot{\Phi}}{a\Phi}\right) 
\label{equationf} 
\end{eqnarray}
\end{subequations}
Here a dot and a prime denote partial derivatives with respect to
$t$ and $y$, respectively. In this work we require that there is
no energy flow from the brane towards the bulk and vice-versa so 
it must be ${}^{(5)}T_t^y = 0$, therefore the choice of the 
functions $F$ and $G$  must give accordingly ${}^{(5)}G_t^y = 0$.
However, as eq. (A.7d)   shows, there is no energy flow 
only for suitable values of the metric coefficients. A 
particularly simple choice which makes ${}^{(5)}G_t^y = 0$ is to 
assume that the metric coefficients in the bulk have the form of 
plane waves propagating in the fifth dimension, so they become 
functions either of the argument $u = t- \lambda\, y$ or of the 
argument $v = t + \lambda\, y$. Of course the particular metric 
which we finally obtain is dependent on this choice. In detail, 
from ${}^{(5)}G_t^y = 0$ we can derive \begin{equation} 
\dfrac{1}{n(u)\,\Phi(u)}\,\dfrac{da(u)}{du} = 
\dfrac{1}{n(v)\,\Phi(v)}\,\dfrac{da(v)}{dv} = \gamma 
\end{equation} where $\gamma$ is a dimensionless constant. Now we
shall assume the $Z_2$ symmetry $y \to -\ y$ and construct a 
solution of eqs. (A.6) by matching a solution depending  only on 
$u$ (for $y>0$) to a solution  depending only on $v$ (for 
$y<0$).\pni The result in (A.8) suggests to multiply  (A.6b) by 
(A.6c) so, taking into account (A.6d), we have 
\begin{equation}
\left[\dfrac{a^2\,(\kappa^2-\varepsilon \, \chi^2)}{A^2\, 
(\kappa^2-\chi^2)}\right]^2\,\left[\left(\dfrac{\partial F}{\partial 
t}\right)\,\left(\dfrac{\partial G}{\partial y}\right) -\, 
\left(\dfrac{\partial F}{\partial y}\right)\, 
\left(\dfrac{\partial G}{\partial t}\right)\right]^2 = 
n^2\,\Phi^2 
\end{equation}
Eliminating $G$ by (A.6a) and taking the square root one finally 
obtains
\begin{equation}
\dfrac{\kappa^2- \varepsilon \, \chi^2}{A\, \kappa^2 \, \chi 
\sqrt{\kappa^2 -\, \chi^2}}\, \left| -\, \left(\dfrac{\partial 
F}{\partial t}\right)\, \left(\dfrac{\partial a}{\partial 
y}\right) + \, \left(\dfrac{\partial F}{\partial y}\right)\,  
\left(\dfrac{\partial a}{\partial t}\right)\right| = n\,\Phi
\end{equation}
Let us first begin working on the $y>0$ side. We put  
$a(t,y)=a(t-\lambda \, y)$ into (A.10) and recalling (A.8)  we 
obtain the following partial differential equation for $F$ 
\begin{equation}
\lambda \, \dfrac{\partial F}{\partial t} +    \dfrac{\partial 
F}{\partial y} = \dfrac{A \kappa^2 \chi \sqrt{\kappa^2 - \chi^2}}
{\gamma \, (\kappa^2 - \varepsilon \, \chi^2)}
\end{equation}
The general solution is
\begin{equation}
F(t,y) = \dfrac{A \kappa^2 \chi \sqrt{\kappa^2 - \chi^2}}
{2 \gamma \lambda \ (\kappa^2 - \varepsilon \, \chi^2)}\, 
(t+\lambda y) + f_{(-)}(t - \lambda \, y)
\end{equation} 
The function $f_{(-)}(u)$ can be determined by (A.6d) 
after eliminating $G$ by (A.6a). The result is
\begin{equation}
\dfrac{df_{(-)}}{du} = \dfrac{A \chi \sqrt{\kappa^2 - 
\chi^2}}{a^2 (\kappa^2-\varepsilon\, \chi^2)}\, 
\left\{\varepsilon\, \dfrac{\chi}{\kappa}\, 
\left(\dfrac{da}{du}\right) + \dfrac{1}{\gamma\, \lambda}\, 
\sqrt{\kappa^4 a^4 + 4 \varepsilon\,\gamma^2 \lambda^2  
\,\left(\dfrac{da}{du}\right)^2}\right\} 
\end{equation}
which can be integrated, once the scale factor $a$ has been 
fixed, reminding that in the limit  $\chi \to 0$  it  must be 
$F(t,y) \to 1$ and so also $f_{(-)}(u) \to 1$. Of course if 
one is only interested  in determining $n$ and $\Phi$ from (A.6b)
and (A.6c) it is sufficient the simple knowledge of the 
derivative of $f_{(-)}(u)$. Obviously $n$ and $\Phi$ will  be a 
function of the scale factor $a$. Proceeding in an analogous 
manner  when working on the $y<0$ side, we obtain 
\begin{equation}
F(t,y) = \dfrac{A \kappa^2 \chi \sqrt{\kappa^2 - \chi^2}}
{2 \gamma \lambda \ (\kappa^2 - \varepsilon \, \chi^2)}\, 
(t-\lambda\, y) + f_{(+)}(t + \lambda \, y)
\end{equation} 
where $ f_{(+)}(v)$ and $f_{(-)}(u)$ are the same function $f$ of
the two different arguments $v$ and $u$. As a consequence we can 
write the function $F(t,y)$ on both sides of the brane at $y=0$ 
simply as
\begin{equation}
F(t,y) = \dfrac{A \kappa^2 \chi \sqrt{\kappa^2 - \chi^2}}
{2 \gamma \lambda \ (\kappa^2 - \varepsilon \, \chi^2)}\, 
(t+\lambda\, |y|) + f(t - \lambda \, |y|)
\end{equation} 
The function $G(t,y)$ can then be easily derived from eq.(A.6a).
Finally, we can obtain from eqs. (A.6b) and (A.6c)  the metric 
coefficients $n(t - \lambda \, |y|)$ and $\Phi(t - \lambda \, 
|y|)$ which are given as a  function of  $a(t-\lambda\,|y|)$ by
\begin{equation}
n^2 = \dfrac{1}{2\,\gamma^2 \lambda^2}\, \left(\kappa^2 a^2 + 
\sqrt{\kappa ^4 a^4 + 4\,\varepsilon\, \gamma^2 \lambda^2\, 
({\overset{\ast}{a}})^2}\right)
\end{equation}
\begin{equation}
\Phi^2 = \dfrac{\varepsilon}{2\,\gamma^2}\, 
\left(-\, \kappa^2 a^2 + \sqrt{\kappa ^4 a^4 + 4\,\varepsilon\, 
\gamma^2 \lambda^2\, ({\overset{\ast}{a}})^2}\right)
\end{equation}                  
where the superscribed asterisk $\overset{\ast}{\frac{}{}}$ denote 
derivative with respect to $w=(t - \lambda \, |y|)$
The bulk line element away from the brane is therefore
\begin{eqnarray}
dS^2 = a^2\, d\sigma_k^2 \hspace{-.3cm}  &&-\, 
\dfrac{1}{2\,\gamma^2 \lambda^2}\, \left(\kappa^2 a^2 + 
\sqrt{\kappa ^4 a^4 + 4\,\varepsilon\, \gamma^2 \lambda^2\, 
({\overset{\ast}{a}})^2}\right)\, dt^2 \nonumber \\
&&+\, \dfrac{1}{2\,\gamma^2}\, 
\left(-\, \kappa^2 a^2 + \sqrt{\kappa ^4 a^4 + 4\,\varepsilon\, 
\gamma^2 \lambda^2\, ({\overset{\ast}{a}})^2}\right)\, dy^2
\end{eqnarray}
as given previously in eq. (16). From eqs. (A.7) we get
\begin{subequations} 
\begin{eqnarray} 
G_r^r = G_\vartheta^\vartheta = G_\varphi^\varphi  = &  
6\, \varepsilon\, \kappa^2 -\,\dfrac{k}{a^2(t - \lambda\, |y|)}
\label{equationa} \\  
G_t^t = G_y^y = & 6\, \varepsilon\, \kappa^2 
-\,\dfrac{3 k}{a^2(t - \lambda\, |y|)} \label{equationb}
\end{eqnarray}
\end{subequations} 
in accordance with the 5D cosmological constant and  the 5D 
energy-momentum tensor given previously in eqs. (19) and (20).
 
\end{document}